%% file: main.tex
\documentclass[sigconf]{acmart}
\input{tools}
\setcopyright{acmcopyright}
\copyrightyear{2021}
\acmYear{2021}
\acmConference[Conference '21]{Conference '21}{July, 2021}{}

\AtBeginDocument{%
  \providecommand\BibTeX{{%
    \normalfont B\kern-0.5em{\scshape i\kern-0.25em b}\kern-0.8em\TeX}}}

\begin{document}

\title{Is a Single Model Enough? \modelname{}: A Multi-Model Ensemble Learning for Semantic Code Search}

\author{Lun Du}
\authornote{Equal Contribution}
\authornote{Corresponding Author}
\email{lun.du@microsoft.com}
\affiliation{
    \institution{Microsoft Research Asia}
    \city{Beijing}
    \country{China}
}

\author{Xiaozhou Shi}
\authornotemark[1]
\authornote{Work performed during the internship at MSRA}
\email{xzh0u.sxz@gmail.com}
\affiliation{%
\institution{Beijing University of Technology}
\city{Beijing}
\country{China}
}

\author{Yanlin Wang}
\email{yanlwang@microsoft.com}
\affiliation{%
\institution{Microsoft Research Asia}
\city{Beijing}
\country{China}
}

\author{Ensheng Shi}
\authornotemark[3]
\email{s1530129650@stu.xjtu.edu.cn}
\affiliation{%
\institution{Xi'an Jiaotong University}
\city{Beijing}
\country{China}
}

\author{Shi Han}
\email{shihan@microsoft.com}
\affiliation{%
\institution{Microsoft Research Asia}
\city{Beijing}
\country{China}
}

\author{Dongmei Zhang}
\email{dongmeiz@microsoft.com}
\affiliation{%
\institution{Microsoft Research Asia}
\city{Beijing}
\country{China}
}

\begin{abstract}
Recently, deep learning methods have become mainstream in code search since they do better at capturing semantic correlations between code snippets and search queries and have promising performance. However, code snippets have diverse information from different dimensions, such as business logic, specific algorithm, and hardware communication, so it is hard for a single code representation module to cover all the perspectives. On the other hand, as a specific query may focus on one or several perspectives, it is difficult for a single query representation module to represent different user intents.
In this paper, we propose \modelname{}, a multi-model ensemble learning architecture for semantic code search. It combines several individual learners, each of which emphasizes a specific perspective of code snippets. We train the individual learners on different datasets which contain different perspectives of code information, and we use a data augmentation strategy to get these different datasets. Then we ensemble the learners to capture comprehensive features of code snippets. 
The experiments show that \modelname{} 
has better results than the existing state-of-the-art methods. 
  
\end{abstract}








\maketitle

\input{src/introduction}
\input{src/model}
\input{src/experiment}
\input{src/results}
\input{src/related_work}
\input{src/conclusion}

\bibliographystyle{ACM-Reference-Format}
\bibliography{ref}

\end{document}

%% file: tools.tex
\usepackage{CJKutf8}
\usepackage{url}
\usepackage{booktabs}
\usepackage{enumitem}
\usepackage{makecell}
\usepackage{hyperref}
\usepackage{xcolor}
\usepackage{graphicx}
\usepackage[export]{adjustbox}
\usepackage{xspace}
\usepackage{multirow}
\usepackage{fancybox}
\usepackage[most]{tcolorbox}
\usepackage{color}

\newcommand{\modelname}{MuCoS}

\newcommand{\Fig}{Fig.\xspace}

\usepackage{listings}
\definecolor{gogetit}{HTML}{3CB371}
\definecolor{gold}{HTML}{FFD700}
\definecolor{redd}{HTML}{FF4D4F}
\definecolor{light-gray}{gray}{0.95}
\definecolor{lightgreen}{rgb}{0.8,1.0,0.8}
\usepackage{caption}
\usepackage[normalem]{ulem}
\newcommand\hl{\bgroup\markoverwith
  {\textcolor{yellow}{\rule[-.5ex]{2pt}{2.5ex}}}\ULon}
 \newcommand\ghl{\bgroup\markoverwith
  {\textcolor{green}{\rule[-.5ex]{2pt}{2.5ex}}}\ULon}

%% file: src/introduction.tex
\section{Introduction}
Code search is the most frequent developer activity in software development process \cite{Caitlin15}. Reusable code examples help improve the efficiency of developers in their developing process \cite{Brandt09, Shuai2020}. Given a natural language query that describes the developer's intent, the goal of code search is to find the most relevant code snippet from a large source code corpus. 

Many code search engines have been developed for code search. They mainly rely on traditional information retrieval (IR) techniques such as keyword matching \cite{Meili15} or a combination of text similarity and Application Program Interface (API) matching \cite{Lv15}. Recently, many works have taken steps to apply deep learning methods \cite{he2016deep,ChoMGBBSB14,wang2019tag2gauss,wang2019tag2vec,yang2020domain} to code search \cite{Gu2018,Cambronero2019,Yan2020,Li2020,Feng2020,Zhu2020,Shuai2020,Ye2020,Haldar2020,Ling2020,Ling2020a,wang2020cocogum}, using neural networks to capture deep and semantic correlations between natural language queries and code snippets, and have achieved promising performance improvements. These methods employ various types of model structures, including sequential models \cite{Gu2018,Cambronero2019,Yan2020,Li2020,Feng2020,Zhu2020,Shuai2020,Ye2020,Haldar2020}, graph models \cite{Ling2020, Guo2020}, and transformers \cite{Feng2020}.

Existing deep learning code search methods mainly use a single model to represent queries and code snippets. However, code may have diverse information from different dimensions, such as business logic, specific algorithm, and hardware communication, making it hard for a single code representation module to cover all the perspectives. On the other hand, as a specific query may focus on several perspectives, it is difficult for a single query representation module to represent different user intents.

\begin{figure}[t]
\begin{tcolorbox}[colback=white,colframe=yellow!50!black,boxrule=0.2mm,bottom = 0pt]
\begin{lstlisting}[language=Java,escapechar=@,linewidth=0.99\columnwidth,xleftmargin=-12pt,frame=single,framesep=0mm,backgroundcolor=\color{white},tabsize=1, caption={},captionpos=b]
public static String replaceHtmlEntities(String @\ghl{content}@, Map<String, Character> @\ghl{map}@) {
    for (Entry<String, Character> @\ghl{entry}@ : escapeStrings.entrySet()) {
      if (@\ghl{content}@.indexOf(@\ghl{entry}@.getKey()) != -1) {
        @\ghl{content}@ = @\ghl{content}@.replace(@\ghl{entry}@.getKey(), String.valueOf(@\ghl{entry}@.getValue()));
      }
    }
    return @\ghl{content}@;
  }
\end{lstlisting}
\end{tcolorbox}
\begin{tcolorbox}[colback=white,colframe=yellow!50!black,boxrule=0.2mm,bottom = 0pt]
\begin{lstlisting}[language=Java,escapechar=@,linewidth=0.99\columnwidth,xleftmargin=-12pt,frame=single,framesep=0mm,backgroundcolor=\color{white},tabsize=1, caption={},captionpos=b]
public static String replaceHtmlEntities(String @\hl{var0}@, Map<String, Character> @\hl{var2}@) {
    for (Entry<String, Character> @\hl{var1}@ : escapeStrings.entrySet()) {
      if (@\hl{var0}@.indexOf(@\hl{var1}@.getKey()) != -1) {
        @\hl{var0}@ = @\hl{var0}@.replace(@\hl{var1}@.getKey(), String.valueOf(@\hl{var1}@.getValue()));
      }
    }
    return @\hl{var0}@;
 }
\end{lstlisting}
\end{tcolorbox}
\caption{Code before and after variable renaming.}
\label{fig:variable_renaming}
\end{figure}

\begin{figure}[t]
\begin{tcolorbox}[colback=white,colframe=yellow!50!black,boxrule=0.2mm,bottom = 0pt]
\begin{lstlisting}[language=Java,escapechar=@,linewidth=0.99\columnwidth,xleftmargin=-12pt,frame=single,framesep=0mm,backgroundcolor=\color{white},tabsize=1, caption={},captionpos=b]
public void doAESEncryption() throws Exception{
		if(!initAESDone) 
			initAES();
		cipher = Cipher.getInstance("AES/CBC/PKCS5Padding");
		//System.out.println(secretKey.getEncoded());
		@\hl{cipher.init(Cipher.ENCRYPT\_MODE, secretKey);}@
		@\ghl{AlgorithmParameters params = cipher.getParameters();}@
		iv = params.getParameterSpec(IvParameterSpec.class).getIV();
		secretCipher = cipher.doFinal(secretPlain);
		clearPlain();
	}
\end{lstlisting}
\end{tcolorbox}

\begin{tcolorbox}[colback=white,colframe=yellow!50!black,boxrule=0.2mm,bottom = 0pt]
\begin{lstlisting}[language=Java,escapechar=@,linewidth=0.99\columnwidth,xleftmargin=-12pt,frame=single,framesep=0mm,backgroundcolor=\color{white},tabsize=1, caption={},captionpos=b]
public void doAESEncryption() throws Exception{
		if(!initAESDone)
			initAES();
		cipher = Cipher.getInstance("AES/CBC/PKCS5Padding");
		//System.out.println(secretKey.getEncoded());
		@\ghl{AlgorithmParameters params = cipher.getParameters();}@
		@\hl{cipher.init(Cipher.ENCRYPT\_MODE, secretKey);}@
		iv = params.getParameterSpec(IvParameterSpec.class).getIV();
		secretCipher = cipher.doFinal(secretPlain);
		clearPlain();
	}
\end{lstlisting}
\end{tcolorbox}
\caption{Code before and after statement permutation.}
\label{fig:permute_statement}
\end{figure}

\begin{figure*}[t]
\includegraphics[width=1.0\textwidth]{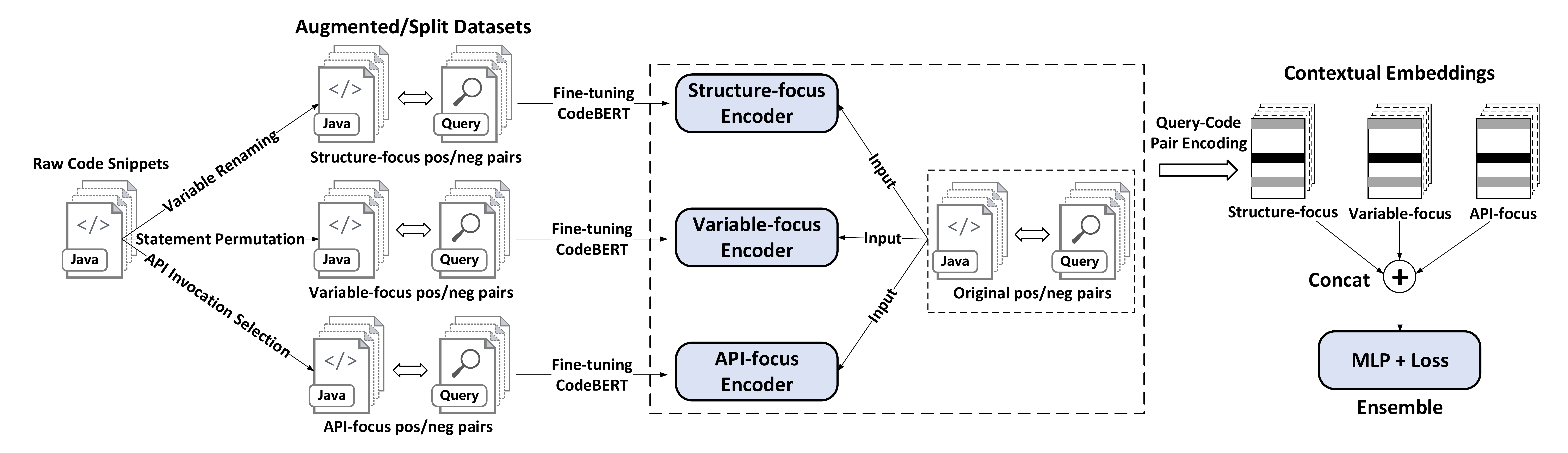}
\caption{An overview of \modelname{}. This framework consists of three phases: data augmentation/separation, individual encoder fine-tuning, and ensemble learning. We first generate three datasets each of which focuses on a specific aspect of code snippets based on data augmentation or separation. Then we learn three individual code search models by fine-tuning three pre-trained CodeBert models based on the generated datasets, respectively. Finally, we use a multi-layer perceptron followed by a concatenation of encodings from three individual models for ensemble learning.} 
\label{fig:overview}
\end{figure*}

To address the problems above, we propose \textbf{\modelname{}}: \textbf{Mu}lti-Model for \textbf{Co}de \textbf{S}earch. First,  we use data augmentation strategy to train multiple models that focus on different perspectives of code. Then, we combine these models using an ensemble learning strategy. For one natural language query in the training corpus, there are multiple corresponding code snippets with the same functionality but different structures or variable names after data augmentation. We believe our model can be inducted to be less focused on the structure or variable information using this data augmentation strategy.

To be specific, we combine three models which focus on structure, local variable, and information of API invocation separately. The three models can roughly represent three typical scenarios in code search where the query is sensitive to business logic, specific algorithm, or hardware communication. 
Since programs have both lexical information and structure information, to build the structure-focused model, we can make the model rely less on local variables so that they will focus more on the structure. In practice, we employ a data augmentation technology, using variable renaming method \cite{Rabin2020} to change the local variable to $varN$ as shown in \Fig~\ref{fig:variable_renaming}, while preserving the original semantics.
%then mix them with the raw data.
Similarly, we learn a variable-focused model based on an augmented dataset where new code snippets are added by permuting two statements of an exiting one if its semantics will not be changed \cite{Rabin2020} as shown in \Fig~\ref{fig:permute_statement}.
%The process to build a local variable focus model is similar, so that we can make our model rely less on structures in programs, and in practice to augment data using statement permutation strategy \cite{Rabin2020} as shown in Figure \ref{fig:permute_statement}, permuting two statements of a program if the semantic of the program will not be changed. 
%These two transformations on programs only change their lexical appearances or syntactical structures, without hurting their original functionality. 
And we also select programs that have API invocations from the JVM library to build a model that focuses on APIs. 
After learning several code retrieval models which represent different characteristics of code, we employ an ensemble strategy to combine the models. Our preliminary results suggest that our method has an improvement of around 12\%  compared to the state-of-the-art baselines.

% We perform data augmentation by generating more code snippets of the invariant semantics but of different lexical appearances or syntactical structures \cite{Rabin2020}, then mix them with raw data. We use our different types of augmented and separated data to train specific models respectively, specifically local variable focus models, structure focus models, and API focus models. These three kinds of models can roughly represent three real scenarios in code search where the query is sensitive to business logic, algorithm and efficiency or hardware communication. After learning several code retrieval models which represent different characteristics of code, we employ an ensemble strategy to combine the models. Our preliminary results suggest that our method have about 12.7\% improvements compared to other state-of-the-art baselines.

% To address the first problem, we perform the data augmentation on code data while keeping the semantics unchanged, as well as combining with pre-training model to further improve the representation ability. For the second problem, we use a multi-layer model structure, each layer builds model with a different behavior, and then combining the different models using xx strategy.

To summarize, this paper makes the following contributions:

\begin{itemize}%[leftmargin=10pt]
% emsemble learning 使得模型可以捕获query intent和code的不同层面的信息
\item We propose a novel multi-model architecture \modelname{} for code search. We use an ensemble learning strategy to capture different perspectives of code information and query intents.

% We use customized data augmentation and data separation strategy to get data segments with different characteristics, and learn different models based on these data, which can represent different real scenarios in code search. To the best of our knowledge, we are the first to employ data augmentation to solve semantic code search problem.
% data augmentation基于程序不变性
\item We do data augmentation based on semantic invariance of programs to get multiple individual learners which focus on different features of code. To the best of our knowledge, we are the first to employ data augmentation to solve the semantic code search problem.

\item We conduct extensive experiments to evaluate the effectiveness of our approach. The results show that our approach significantly outperforms the  state-of-the-art methods by 12\% on the standard dataset and 14\% on the sampled dataset.
%by at least 12\%, and on the sampled dataset, this number can be 14\%. 
% And with ablation study and case study, we provide strong evidence that \modelname{} leads to better understanding of specific aspects of code.
\end{itemize}

% The rest of this paper is organized as follows: \Sec~\ref{sec:model} introduces the design of \modelname{}. We present details about our experiment settings in
% \Sec~\ref{sec:exp}. We compare \modelname{} with state-of-the-art works, conduct ablation study and case study in \Sec~\ref{sec:eval}. Sec~\ref{sec:rw} describes the related work. Finally, we conclude our work in \Sec~\ref{sec:con}.

%% file: src/model.tex
\section{Proposed Model: \modelname{}}\label{sec:model}
\Fig~\ref{fig:overview} illustrates the overall structure of the proposed model framework \modelname{}. This framework consists of three phases: data augmentation/separation, individual encoder fine-tuning, and ensemble learning.

The basic idea of our method is to learn several individual encoders, each of which focuses on a specific aspect of code snippets based on different datasets. These datasets are generated from the original dataset through domain knowledge of programming language, such as API information from JVM library and semantic equivalent transformations of code snippets \cite{Rabin2020}. Finally, we leverage ensemble learning to integrate the individual modules.

 % In this model framework, each model's architecture follows the architecture of CodeBERT \cite{Feng2020}. Then we use our augmented and separated data to fine-tune the pre-trained models, producing different models which are agnostic to specific characteristics. Finally, we use ensemble learning strategy to combine the models.
 
% \subsection{Model Architecture}
% We follow CodeBERT \cite{Feng2020} to use use multi-layer bidirectional Transformer as the model architecture, which is the same as RoBERTa-base. It has 12 layers, and each layer has 12 self-attention heads, and the size of each head is 64. The hidden dimension is 768 and the inner hidden size of the feed-forward layer is 3072. The total number of model parameters is 125M.

\subsection{Data Augmentation or Separation}
% \origin{In this step, we design three data generation or separation strategies for the building of three kinds of models: local variable focus data generation, structure focus data generation, and API focus data generation. The three data generation strategies are listed as follow:}

The first step is to generate proper datasets for feeding individual learners. We design three data augmentation or separation strategies for separately building a structure-focused dataset, a variable-focused dataset, and an API-focused dataset: % when the natural language query is not sensitive to business logic, algorithm and efficiency or hardware communication.
\begin{itemize}
    \item \textbf{Structure-focused data generation:} To reduce the impact of local variable names and make the model focus more on the program structure, we use the variable renaming program transformation method in \cite{Rabin2020} to generate semantic equivalent code snippets with local variable names replaced by $varN$, as shown in \Fig~\ref{fig:permute_statement}. Then we mix them with origin data.
    \item \textbf{Variable-focused data generation:} To reduce the impact of structure and make the model focus more on lexical information, we use the statement permutation program transformation method in \cite{Rabin2020} to change the program structure, swapping two independent statements (i.e., with no data or control dependency) in a basic block of a method while maintaining semantic equivalence, as shown in \Fig~\ref{fig:variable_renaming}. Then we mix them with origin data.
    \item \textbf{API-focused data generation:} We select the samples that the code snippets has API invocation from JVM library as the training data for API-focused model. 
\end{itemize}
The general idea behind the first generation method is that the model will not highly rely on variable names to determine the similarity between query-code pairs, because the code snippets with different variable names will correspond to the same query, and hence the model will pay more attention to structural characteristics. Similarly, the model fine-tuned by the second dataset will pay more attention to variable names of code snippets. The generation method of the third one is easy to understand via directly selecting the code snippets with API invocation.

% For the first two data generation strategies, we mix the generated code snippets with the origin as the fine-tuning data. For the last strategy, we only separate the data and use the part with API sequences. 
% We will further use the separated two kinds of data which has API sequences or without API sequences to train two models and ensemble them as our API agnostic model. 
% Many programming specifications require software engineers to use meaningful local variable names to make the programs they write more readable and maintainable, especially when they are writing code about business logic, since in this scenario, the program logic is tend to be straightforward but have strong connection with specific function module. 

% To capture the different characteristics from the perspective of query, we embed the queries in our dataset using BERT, and use K-means algorithm clustering them in $N$ clusters, which can roughly represent data with different kinds of queries. % need a reason

\subsection{Individual Model Fine-Tuning and Model Combination}
We follow CodeBERT \cite{Feng2020} to use a multi-layer bidirectional Transformer as the model architecture.
%which is the same as RoBERTa-base. 
% It has 12 layers, and each layer has 12 self-attention heads, and the size of each head is 64. The hidden dimension is 768 and the inner hidden size of the feed-forward layer is 3072. The total number of model parameters is 125M.
% To organize the data, we set the input as the concatenation of code snippet and natural language queries with a special separator token. Following the standard way of processing text in Transformer, we regard a natural language text as a sequence of words, using WordPiece \cite{Wu2016} to split it, and regard a piece of code as a sequence of tokens. Then we feed input like that into the base model to fine-tune it.
We feed positive and negative samples in the model, and the ratio of them is 1:1. To build negative examples, for each positive sample we change the query into a randomly mismatched one while the code snippet remains unchanged. 

% \subsection{Model Selector}
% After training some models that represent different characteristics of code or queries, we need to make a selection to choose several appropriate models for each code search activity. Here we introduce a selection strategy based on attention mechanism as shown in Figure 2. We compute the attention between the embedding from our models and the embedding from the origin pre-trained models. Given a set of code embedding vectors $\{e_1, e_2, ..., e_n\}$, the attention weight $\alpha_i$ for each $e_i$ can be computed as:
% $$
% \alpha_{i}=\frac{\exp \left(a_{c} \cdot e_{i}^{\top}\right)}{\sum_{i=1}^{n} \exp \left(a_{c} \cdot e_{i}^{\top}\right)}
% $$
% Then we can compute the final embedding vector $e$ as:
% $$
% e=\sum_{i=1}^{n} \alpha_{i} e_{i}
% $$

%\subsection{Model Combination}
We use an ensemble learning strategy to combine the structure-focused model, variable-focused model, and API-focused model. We first concatenate the embedding of the hidden state's last layers of these models, then append an MLP classifier with two linear layers to the last layer of the concatenated neural network, training and updating its weights using the origin data. The loss functions of individual model fine-tuning and ensemble learning are both cross entropy to discriminate the positive pair and the negative pair.
%\begin{equation}
%
%\end{equation}

%After this ensemble learning phase, we get our final combination model.

%% file: src/experiment.tex
\section{Experimental Setup}
\label{sec:exp}
\subsection{Dataset}
To evaluate the effectiveness of our method, we use a widely used dataset for the code search task: CodeSearchNet \cite{Husain2019}, containing $500, 754$ pairs of function-level Java code snippets and their descriptions. There are $454, 443$ pairs as the training set, $30, 655$ pairs as the validation set and $26, 909$ pairs as the test set. 

We extend the training dataset by collecting the same number of negative samples as the positive samples. For each positive pair sample, we randomly select a mismatched query replacing the original query while maintaining the code snippet unchanged to construct a negative pair sample.

\subsection{Evaluation Metrics}
We use three widely used metrics for the evaluation of code search methods: 
FRank, SuccessRate@k, and MRR.
The FRank, or best hit rank, is the rank of the first hit result in this result list. A smaller FRank implies lower inspection effort for finding the desired result \cite{Gu2018}. FRank can represent the effectiveness of a single code search query.
The $SuccessRate@k$ measures the percentage of queries for which more than one correct result exists in the top k ranked results. In our evaluation, is calculated as follows:
$$
    \text { SuccessRate@k }=\frac{1}{|Q|} \sum_{q=1}^{Q} \delta\left(\text { FRank }_{q} \leq k\right)
$$
where $Q$ is a set of queries, $\delta\left(\cdot\right)$ is a characteristic function, i.e., $\delta\left(\cdot\right) = 1$ if $\cdot$ satisfied, otherwise $\delta\left(\cdot\right) = 0$. A higher $SuccessRate@k$ means better code search performance. The MRR is the average of the reciprocal ranks of results of a set of queries $Q$. The reciprocal rank of a query is the inverse of the rank of the FRank. In our evaluation, is calculated as follows:
$$
MRR=\frac{1}{|Q|} \sum_{q=1}^{|Q|} \frac{1}{FRank_{q}}.
$$

%% file: src/results.tex
\section{Results}\label{sec:eval}
In this section, we present the results of our experiments and answer three research questions. We also provide a case study.

\subsection{RQ1. Does \modelname{} outperform SOTA deep code search methods?}
%\textbf{Does \modelname{} outperform the state-of-the-art deep code search methods?}

Table \ref{tab:baseline} shows the evaluation results of \modelname{} compared to several state-of-the-art deep code search models: five baseline models provided by CodeSearchNet \cite{Husain2019}, the classical joint embedding model CODEnn \cite{Gu2018}, and the pre-trained model CodeBert \cite{Feng2020}. Among the SOTA models, \modelname{} achieves the best performance, with MRR 12\% higher than the best baseline. We test our method on five seeds, the variance is of the order of 1e-5, which can be ignored and shows that our method is stable.

% we lack of good results of baselines
% \begin{table}[h]
% \centering
% \caption{Evaluation of different baselines on full Code Search Net corpus. The results in this table are average results on five random seeds. The results in this table are average results on five random seeds. T represents that we including training while ensemble the models.}
% \begin{tabular}{l|cccc}
% \hline
% Baseline      & S@1     & S@5 & S@10 & MRR\\ \hline                 
% CodeSearchNet &      &     &     & 0.588 \\ \hline
% CODEnn        &      &     &     & 0.60\\ \hline
% CodeBert      &      &     &     & 0.708\\ \hline
% CQIL          &      &     &     & 0.54\\ \hline
% CARLCS-CNN    &      &     &     & 0 \\ \hline % not necessary
% \modelname{}(T)    &      &     &     & 0.787 \\ \hline

% \end{tabular}
% \label{tab:baseline}
% \end{table}

\begin{table}[t]
\centering
\caption{Evaluation of different baselines on full Code Search Net corpus.}
\begin{tabular}{l|cccc}
\hline \bottomrule 
Model      & S@1     & S@5 & S@10 & MRR\\\hline             
NBoW &0.499      & 0.698   & 0.752    & 0.589 \\ \hline
1D-CNN & 0.424  &   0.631  &  0.699   &  0.518\\ \hline
biRNN &  0.485  &  0.685   &  0.743 & 0.644\\ \hline
SelfAtt &  0.486  &  0.682  &  0.738 &  0.575\\ \hline
ConvSelfAtt &  0.413  &  0.619  &  0.681 &  0.507\\ \hline
CODEnn         &0.146  &  0.146 & 0.146  & 0.146\\ \hline
CodeBert      & 0.642 &  0.792   &  0.825 & 0.708\\ \hline
% CQIL         &0.4452+     &0.6465+     & 0.7148+    & 0.53+\\ \hline
% CQIL         &0.464     &0.659     & 0.723    & 0.547\\ \hline
% CARLCS-CNN    &      &     &     & 0 \\ \hline % not necessary
% \modelname{}-C    &   0.716  &  0.834  &  0.854 & 0.770\\ \hline % on DXG 112
\modelname{}  &  \textbf{0.750}  &  \textbf{0.843}  &  \textbf{0.860}  & \textbf{0.793} \\  \hline
\bottomrule

\end{tabular}
\label{tab:baseline}
\end{table}

\subsection{RQ2. How does \modelname{} perform compared to other baselines on a small dataset?}
%\textbf{How does our method performed compared to other baselines on a small dataset?}

We try a small sampled dataset with 60k samples to test whether code search models can maintain good performance on a small dataset. The training set is sampled from CodeSearchNet's training set, the number of training data is 60000, while the validation data and testing data keep the same amount. The baselines are trained with default parameters. Table \ref{tab:sampled} shows that the performance of most baselines drop largely on the small dataset. Our method and CodeBert have significantly less performance drop than other methods. We assume that pre-trained models can better adapt to scenarios with small data on the code search task. Moreover, our method still has a 14\% advantage compared to CodeBert on the small dataset.
% \begin{table}[h]
% \centering
% \caption{Evaluation of different baselines on CodeSearchNet. The results in this table are average results on five random seeds. T represents that we including training while ensemble the models.}
% \begin{tabular}{l|cccc}
% \hline
% Baseline      & S@1     & S@5 & S@10 & MRR\\ \hline
% NBoW &      &     &     & 0.51 \\ \hline
% CODEnn        &      &     &     & 0.60\\ \hline
% CodeBert      &      &     &     & 0.708\\ \hline
% CQIL          &      &     &     & 0.54\\ \hline
% CARLCS-CNN    &      &     &     & 0 \\ \hline % not necessary
% \modelname{}(T)        &      &     &     & 0.787 \\ \hline

% \end{tabular}
% \label{tab:sampled}
% \end{table}
\begin{table}[!t]
\centering
\caption{Evaluation of different baselines on sampled Code Search Net corpus.}
\begin{tabular}{l|cccc}
\hline \bottomrule 
Model      & S@1     & S@5 & S@10 & MRR\\ \hline
NBoW & 0.271     &0.441     &0.507     & 0.354 \\ \hline
1D-CNN & 0.052  &0.151  & 	0.224 &  0.110\\ \hline
biRNN &  0.178  &  0.364   & 0.454 &  0.270 \\ \hline
SelfAtt & 0.298 & 0.487  & 0.562 & 0.388 \\ \hline
ConvSelfAtt & 0.193 & 0.375  & 0.461 & 0.282 \\ \hline
CODEnn        &0.043     &0.043     &0.043     &0.043\\ \hline
CodeBert      & 0.624  &  0.773  &  0.805   & 0.661\\ \hline % wait future check
% CQIL          & 0.448 & 0.666 & 0.751 & 0.541 \\ \hline % previous MRR: 0.486
% CARLCS-CNN    &      &     &     & 0 \\ \hline % not necessary
\modelname{}  & \textbf{0.702}  & \textbf{0.815} & \textbf{0.831} & \textbf{0.754} \\ \hline
\bottomrule  

\end{tabular}
\label{tab:sampled}
% \vspace{-20pt}
\end{table}

\subsection{RQ3. How the individual models in \modelname{} affect its overall effectiveness?}
\begin{figure}[t]
\centering
\includegraphics[width=0.45\textwidth]{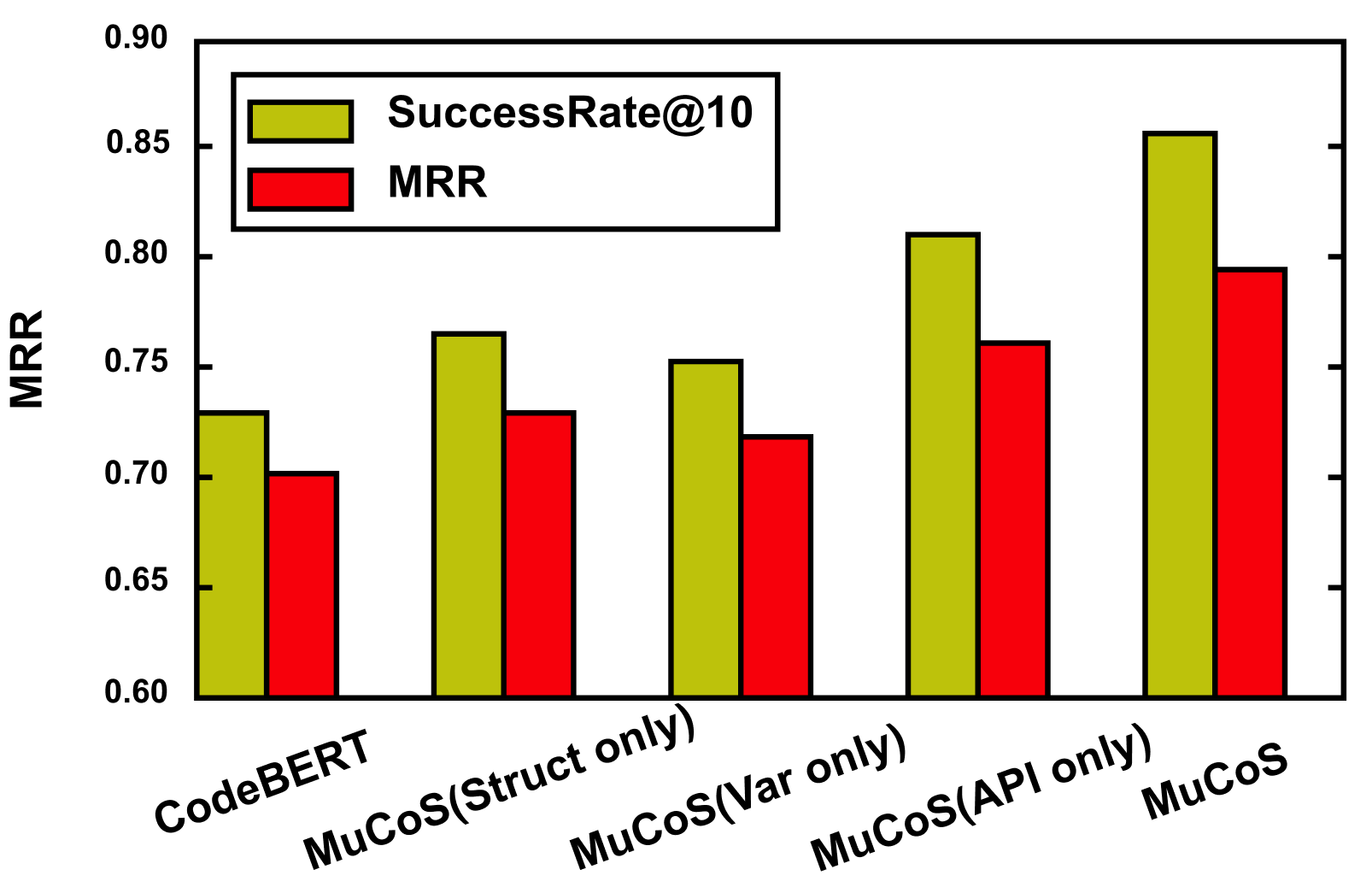}
\vspace{5pt}
\caption{Evaluation of individual model in \modelname{}. } 
\label{fig:individual}
% \vspace{-6pt}
\end{figure}

% add a 柱状图 
In this research question, we evaluate whether each individual model contribute to building our final model \modelname{}. 
%Besides the evaluation for our resulting model, we also evaluate every intermediate model which contributes to building our final model. 
\Fig~\ref{fig:individual} shows the results. Surprisingly, the three models that capture individual features all perform better than the baselines. The reason may be that the baselines do not model information of different features separately, %which causes the baselines to be interfered by the information of different features. 
which causes the baseline model to be confused with information from different features. 
On the other hand, \modelname{} %the effect of ensemble 
is significantly better than the three individual learners, which is also in line with the basic assumptions of ensemble learning.
% \begin{table}[h]
% \centering
% \caption{Evaluation of individual model in \modelname{}. For our \modelname{} method, T represents for ensemble model with training, F represent full dataset, while S represent sampled dataset. The results in this table are average results on five random seeds. }
% \begin{tabular}{l|c}
% \hline
% Baseline      & MRR\\ \hline
% CodeBert      &  0.708\\ \hline
% variable agnostic &     0.714\\ \hline
% structure agnostic &      0.725\\ \hline
% exist api &    0.764\\ \hline
% w/o api &      0.749\\ \hline
% api agnostic &      0.795\\ \hline
% code model &       0.750\\ \hline
% code model(T) &       0.7743\\ \hline
% \modelname{}(F)     &      0.7714\\ \hline
% \modelname{}(T+F)    &   0.7814 \\ \hline
% \end{tabular}
% \label{tab:tmp}
% \end{table}

\subsection{Case Study}
We now provide an example to show our individual models can capture the specific feature and contribute to the performance of \modelname{}. 
\Fig~\ref{fig:3-1} shows the corresponding code snippet for the natural language query "get the field label". We evaluate them on our individual models and find that the correct result ranks first in the API-focused model, third in the structure-focused model, fifth in the var-focused model, and also first in our \modelname{} model. 

In this case, the API-focused model can better capture code features than the structure-focused model and the var-focused model. Since the query has many overlaps with the APIs in the code snippet, the API-focused model can capture more information and have better performance. It also contributes to the overall performance of \modelname{}.

\begin{figure}[t]
\begin{tcolorbox}[colback=white,colframe=yellow!50!black,boxrule=0.2mm,bottom = 0pt]
\begin{lstlisting}[language=Java,escapechar=@,linewidth=0.99\columnwidth,xleftmargin=-12pt,frame=single,framesep=0mm,backgroundcolor=\color{white},tabsize=1, caption={},captionpos=b]
public static String createLabelWithNameSpace( final String namespace, final String fieldName, final ResourceBundle bundle ) {
      String label;
      try { label = bundle.getString( namespace + '.' + fieldName );
      } catch ( MissingResourceException mre ) {
          label = generateLabelValue( fieldName ); }
      return label;
}
\end{lstlisting}
\end{tcolorbox}
\caption{The corresponding code snippet to the query "get the field label".}
\label{fig:3-1}
\end{figure}

%% file: src/related_work.tex
\section{Related Work}\label{sec:rw}
% code search 
Code search is a cross-field of natural language processing and software engineering which aims to retrieve code snippets from a large code corpus that most match the developer's thoughts using natural language. There are mainly two kinds of approaches of code search, information retrieval (IR) based \cite{Meili15, Lv15} and deep learning based \cite{Gu2018,Cambronero2019,Yan2020,Li2020,Feng2020,Zhu2020,Shuai2020,Ye2020,Haldar2020,Ling2020,Ling2020a}.

% IR based
Most of the existing code search engines rely on IR-based techniques, employing keyword matching or text similarity to retrieve code snippets. 
% For example, \cite{Meili15} expanded a query with synonyms from WordNet and performed a keyword matching. \cite{Lv15} proposed CodeHow, which combines text similarity and API matching through an extended Boolean model.
% deep learning code search
Recently, deep learning methods become mainstream since they do better at capturing deep and semantic correlations between code snippets and search queries and have promising performance. For example, \cite{Gu2018} proposed CODEnn, which can jointly embed code snippets and natural language descriptions into a high-dimensional vector space and then compute similarity. \cite{Feng2020} present CodeBERT, a bimodal pre-trained model for natural language and programming language which can solve code search problem.
% For example, \cite{Gu2018} proposed CODEnn, which can jointly embed code snippets and natural language descriptions into a high-dimensional vector space, then code snippets semantically related to a natural language query can be retrieved according to their vectors. \cite{Li2020,Zhu2020,Haldar2020} delivered several methods to capture the internal semantic correlations between code snippets and queries. \cite{Feng2020} present CodeBERT, a bimodal pre-trained model for natural language and programming language, which can capture the semantic connection between them and produces general-purpose representations for code search.

%% file: src/conclusion.tex
\section{Conclusion}\label{sec:con}
In this paper, we propose a novel multi-model architecture \modelname{} for code search. Instead of training one single model for code snippets and queries, we train multiple models which have specific features and combine them, which can help us better capture the diverse meaning from code snippets and natural language queries. To train models with specific features, we use a data augmentation and separation strategy to force the models to capture features of specific perspectives. Then we use an ensemble learning strategy to combine our models. Our experimental study has shown that the proposed approach is effective and outperforms other state-of-the-art approaches.

This work is ongoing. In the future, we plan to evaluate our methods on more datasets. And we will train more individual learners focusing on more features. Moreover, we want to build a selection module to help select individual learners which are most suitable for the search query.